# Temperature Study of Sub-Micrometric ICs by Scanning Thermal Microscopy

S. Gomès, O. Chapuis, F. Nepveu,  N. Trannoy, S. Volz, B. Charlot, G. Tessier, S. Dilhaire, B. Cretin and P. Vairac

*Abstract*—**Surface temperature measurements were performed with a Scanning Thermal Microscope mounted with a thermoresistive wire probe of micrometric size. A CMOS device was designed with arrays of resistive lines 0.35µm in width.  The array periods are  0.8  µm and 10µm to study the spatial resolution of the SThM. Integrated Circuits with passivation layers of micrometric and nanometric thicknesses were tested. To enhance signal-to-noise ratio, the resistive lines were heated with an AC current. The passivation layer of nanometric thickness allows us to distinguish the lines when the array period is 10µm. The results raise the difficulties of the SThM measurement due to the design and the topography of ICs on one hand and the size of the thermal probe on the other hand.**

## I. INTRODUCTION

Semiconductor devices are continuously reduced in size for integration. They must withstand ever higher electric fields which induce severe degradation or total failure. Measurements of  power sources with sub-µm scales have then become fundamental for microelectronics.

While far field optical methods such as laser reflectance thermometry and micro-Raman thermometry are limited by diffraction, the Scanning Thermal Microscope (SThM) is a relatively new technique allowing for such investigations [1]. The principle of the method is based on the control of the thermal near field interaction between a small thermal probe and a surface. The technique uses Atomic Force Microscopy (AFM) to keep the probe-sample interatomic force constant and to accurately scan the surface when imaging is performed. Its key element is the thermal probe. For fifteen years, various thermal probe prototypes (thermocouple or thermoresistive elements integrated at the apex of AFM probes) [1] have been proposed. We use the thermoresistive wire probe developed by Dinwiddie et al in 1994 [2]. Thermal heat transfers [3-6] and materials thermophysical properties [7] were analysed with a submicrometric spatial resolution when the probe is heated by Joule effect . However its ability to measure quantitative temperatures on ICs at submicrometric scale are not yet proven. Fiege et al. [8] have demonstrated that this tool can allow for the localization of failures in integrated devices at saturation or in breakdown condition but on larger scales. The SThM has also been used to investigate the mechanisms of electric degradation inside a metal-oxide-silicon (MOS) structure submitted to an increasing electric stress [9]. More recently, quantitative temperature variations on PN thermoelectric couples [10] were measured.

The spatial resolution of scanning probes is determined by the characteristic length of the predominant physical mechanism governing the probe-sample interaction (and then implicitly by the probe size). In the case of the thermoresistive probe, Figure 1 illustrates the heat transfers operating between the probe and a sample:  solid-solid conduction, conduction via the water meniscus and through the surrounding gas and near field radiation. Solid-solid contact and water meniscus with averaged radii 20-50 nm and 100-200 nm respectively [3-6] have been shown to favor the tip-sample heat transfer [2,4-6]. The detection of submicrometric area can then be expected in spite of the micrometric size of the considered tip. What is the real ability of the SThM in terms of spatial resolution and sensitivity, to detect a local heat dissipation on ICs devices ?

This work tackles the problem of the SThM response on active integrated structures by modeling and measuring several ICs configurations. A dedicated CMOS device was designed with arrays of resistive lines with periods of 0.8   µm and 10µm. The resistive lines are 0.35µm in width. ICs covered by passivation layers with thicknesses of 3,6µm and  of a few nanometers were tested. In the last case, the topography of the arrays under the layer was shown. After the presentation of the experimental technique in the following part, the description of the samples and their used conditions are given and  the modeling used to predict the thermal behavior of the component with thick passivation layers is presented. The results of the measurements with the SThM are next addressed and analyzed in section IV. Results obtained for a sample with the thinner passivation layer are also discussed. Finally, some concluding remarks and recommendations are given in section V.

Manuscript received January 23, 2006. This work was supported in part by the French  National Centre of  Scientific Research (CNRS).

S. Gomès is with the Centre de Thermique de Lyon, UMR CNRS 5008, Institut National des Sciences Appliquées de Lyon, Université Claude Bernard Lyon 1, 20 Avenue Albert Einstein, 69621-F Villeurbanne cedex, France (phone: 334-7243-6428 ; fax: 334-7243-8811; e-mail: severine.gomes@insa-lyon.fr). As the other authors of this paper, she is member of the French CNRS Group (Groupement de Recherches CNRS 2503) called "Heat Transfer at Micro- and Nano-scales", coordinated by S. Volz (Groupement de Recherche Micro et Nanothermique, GDR CNRS 2503, Ecole Centrale Paris, Grande Voie des Vignes, 92295 Châtenay Malabry, France ; phone: 331-4113-1049; e-mail: volz@em2c.ecp.fr).

## II. EXPERIMENTAL SET UP

The experimental set up used in this study results from the combination of a SThM working in the DC mode and a sample submitted to an AC electric current (see Figure 2).

The SThM is a scanning probe microscope marketed by Veeco Corporation. As introduced before, the thermal probe is a thermoresistive element consisting in a platinum wire 5μm in diameter and 200μm in length. The wire is bent to form a loop [2]. The thermal sensor is used as a thermometer. It constitutes one of the legs of a Wheastone bridge. As the temperature of the probe varies, the corresponding change in the probe electric resistance leads to a modification of the voltage balance of the bridge and then of the output voltage ($V_{AB}$) of the circuit. The control and the amplification of the output voltage during the scan allow for generating the thermal image. The contrast of this image corresponds to the variations of the probe electric resistance and the temperature averaged on the length of the platinum wire. Those variations are due to modifications of the temperature of the sample surface. This contrast only provides qualitative information about the temperature field. The calibration of the technique is required for quantitative measurements. There actually exists no general calibration protocol because the measure depends on the tip-sample contact as well as on the geometry and thermal properties of the sample. The calibration of the technique constitutes one of the objectives of our further works. The thermal control unit being independent of the AFM system, topographical and thermal images can be obtained simultaneously.

When studying a resistive sample, a second probe is put on the electric contact of the considered specimen to inject a current in the sample to heat it by Joule effect. The sample can be submitted to a sinusoidal current $I$ at a frequency $f = \omega/(2\pi) : I = I_0 \sin(\omega_s t)$. In this configuration, the internal heating of resistive elements is proportional with $I^2 = 0.5 I_0^2 (1 + \sin(2\omega t - \pi/2))$ and occurs at $f_s = 2f$. The contrast of SThM images is obtained by measuring the amplitude and the phase of $V_{AB}$ at the $2f$ harmonic by using a Lock-In Amplifier (LIA).

Let us notice that because the thermal behavior of the probe follows a low-pass filter law and because the time response in air is found to be about 200-300μs [11], a.c. operations at low frequencies are usually required to enhance the SThM sensitivity (<800Hz). In the experiments, presented in section IV of this paper, the frequencies of the thermal excitation of samples were chosen to be lower than the frequency cut-off of the tip so that the thermal response of probe follows the thermal response of the sample surface with no dephasing compared to the excitation.

## III. THE ICS SAMPLES

### A. Samples Design

The sub-micrometric ICs investigated consist in 0.35μm in width polysilicone strips resistors. As shown in Figure 3, the studied structures are composed with nine resistors in parallel and spaced of 10μm or 0.8μm. Resistors are insulated from their polysilicone substrate by a $SiO_2$ layer of about 300 nm in thickness (gate and field oxides in Figure 4). In a first and original configuration ('samples 1'), the resistors are covered by a 3.6μm in thickness $SiO_2$ passive layer. To obtain a second configuration ('samples 2'), the passivation layer of the original configuration has been chemically etched so that a thin layer of some nanometers in thickness covers the resistors (spaced of 10μm only in this case). Table 1 synthesizes these data and gives the topography aspect for each of the investigated samples .

In order to deal with the samples surface, a study of the topography of samples were performed by AFM. Figure 5 and Figure 6 give AFM images and topographical profiles representative of results obtained on the surface of samples of type 'samples 1' and 'samples 2' respectively. Very small features of a few nanometers in height that are the signature of the resistors locations under the 3.6μm in thickness passivation layer were measured on 'samples 1'. Given that so small topographical features are negligible facing the roughness of the micrometric probe, 'samples 1' were used to model flat chips including buried heating sources.

In Figure 6, large and deep holes are placed in a prominent position at the strips locations. 'Samples 2' were then used to simulate components with surface heating sources and topography.

### B. Samples Heating Conditions

In all the experiments given in this paper, the resistive samples were submitted to a sinusoidal current (see in Table 2) to allow for SThM measurements in the ac regime. In the ac regime, LIA indeed provides a bandwidth reduction which yields higher signal-to-noise ratio than dc measurements.

### C. Thermal Model

In order to predict the thermal response of the SThM for measurements on the samples of type 'samples 1', a model was created to simulate the a.c. thermal behavior of these samples. In this model, the geometry of the heating sources is assumed of such a kind that a simplification into a two-dimensional problem can be performed (case of an infinite wires array). As shown in Figure 7, sample is described as a multiplayer constituted up to down of an oxide layer and a polysilicon substrate.

In ac regime, the heat propagation in each layer, indexed 'n' ($n = 1$ for the oxide and 2 for the substrate), is described by the harmonic Fourier equation:

$$\Delta T_n(x,y,z) + i\frac{\omega}{\alpha_n} T_n(x,y,z) = \frac{-S_n(x,y,z)}{K_n} \qquad (1)$$

where the following $T_n(x,y,z) = T_n(x,y,z).\exp(-i\omega t)$ phase convention has been retained. In this equation, $T_n(x,y,z)$ is the temperature a.c. component; $x$, $y$, $z$ are the space coordinates; $t$ and $\omega$ are respectively the time and the time angular frequency of the heating source ($\omega=2\pi f_s$); $\alpha_n$ and $K_n$ are respectively the thermal diffusivity and conductivity of the layer $n$ and $S_n$ is the power density of the source in the layer $n$.

According with the configuration of our samples, $S_2=0$ in the substrate. In the oxide layer, the heat sources distribution (nine heating lines spaced by a period of 0.8 μm and 10μm) is modeled by describing each heating line section as a limited flat heat source with power density P. This heating sources distribution generates from its plane (in $z=z_e$ in Figure 7) the thermal waves in the whole system. In our approach, the induced thermal waves satisfy the following conditions:
- Continuity of both temperature and heat flow at the interface between the oxide and polysilicon substrate (in $z = z_S$ in Figure 7).
- A.C. thermal bath ($T(x,y,z=e)=0$ at the back of the sample).
- Adiabatic sample surface (in $z=0$). This last assumption is supported by the fact that the convective and radiative heat transfer coefficient of a hot horizontal slab is less than 10W.m$^{-2}$K$^{-1}$. This includes linearized radiation. The 1D heat transfer coefficient of the material between the resistances and the surface is thermal conductivity divided by thickness: $1.28/(3,6.10^{-6})$=$3,6.10^5$W.m$^{-2}$K$^{-1}$. The heat flux will be reduced $3,6.10^4$ times when crossing the boundary. The boundary limit can therefore be assumed as adiabatic to describe the temperature field in the oxide above the resistances.

In each homogeneous layer of our sample description, the a.c. temperature field may be expanded as a double plane wave spectrum [12]:

$$T_n(x,y,z) = \frac{1}{2}\int_{-\infty}^{+\infty}dk_x\int_{-\infty}^{+\infty}dk_y(\tilde{T}_n^+(k_x,k_y)\exp(i\chi_n z)$$
$$+\tilde{T}_n^-(k_x,k_y)\exp(-i\chi_n z) \tag{2}$$

with

$$\chi_n = \left(\frac{i\omega}{\alpha_n}-k_x^2-k_y^2\right)^{1/2} \qquad \mathrm{Re}\,(\chi_n)>0 \tag{3}$$

and where $k_x$ and $k_y$ are the Fourier space coordinates; $\widetilde{T_n^+}$ and $\widetilde{T_n^-}$ are respectively the forward and the backward components of the 2D Fourier transform of the ac temperature field in the layer $n$. Each spectrum is related with the others by the boundary conditions given above. The development of these boundary conditions in the Fourier space then permits to precise the a.c. temperature field in each part of the system. At the sample surface, i.e. in $z = 0$, the 2D Fourier transform of the ac temperature field can be written as:

$$\tilde{T}_1(k_x,k_y,0) = \frac{i\tilde{S}(k_x,k_y)}{2K_1\chi_1}\exp(-i\chi_1 z_e)(1+B) \tag{4}$$

with

$$B = \frac{1-A}{1+A}\exp(i\chi_1 2z_s) \tag{5}$$

where

$$A = \frac{K_2\chi_2}{K_1\chi_1}\frac{\exp(i\chi_2 z_s)+\exp(i\chi_2(2e-z_s))}{\exp(i\chi_2 z_s)-\exp(i\chi_2(2e-z_s))} \tag{6}$$

and $\tilde{S}(k_x,k_y) = FFT(S(x,y))$ is the 2D Fourier transform of the ac heat source in $z=z_e$.

Equations (2) and (6) give access to the ac temperature field on the sample surface. In all the numerical simulations of this temperature field, given in this paper, we used 2D fast Fourier transform (FFT) algorithms (a bounded window and a bounded number of points; the 2D areas were limited here to 1200 X 1200 points). We here used a Gaussian distribution of the heating power for each of the limited flat heat sources describing the heating lines. This Gaussian approximation was made with an effective radius $r$ at 1/e equal to the half of the line width ($r$=175mm) and an integrated value of power which is the double of the experimental one integrated on a cube of area $2r \times 2r \times b$ where $b$ is the polysicon P1 resistor thickness ($b$=300nm). As in the experiments presented in this paper, the modulation frequency of heat sources was set to 140Hz in our calculi. In Table 3 are listed thermophysical properties and geometrical characteristics involved in the samples description.

Figure 8 gives the simulated amplitude of the a.c. temperature field at the surface of the sample with the resistors spaced of 0.8μm and heated by an input power of 70mW. It is clearly shown that the wires can not be thermally revealed by probing the sample surface. This is due to the heat spreading from the resistors in the SiO$_2$ layer which is too thick compared to the space between the heat sources.

Similar simulations were made for the structure with resistors spaced of 10μm under a power of 275mW. Results are given in Figure 9. In this case, the distance between the resistive lines appears large enough so that the a.c. temperature profile at the surface indicates the contribution of each resistors.

## IV. STₕM MEASUREMENTS, RESULTS AND DISCUSSION

The experimental set-up presented in section II was used to investigate the a.c. component of the temperature field on the surface of each sample described in section III. Measurements were performed in ambient.

### A. 'Samples 1': With Thick Passivation Layer

Figure 10 reports the thermal contrast in amplitude obtained at 140Hz and the thermal signal corresponding with two scan lines drawn in the middle of the structure surface and along but perpendicularly to the resistors lines for the wire array with a 0.8μm period. It clearly appears that the wires cannot be distinguished. As found by simulation with the model in the section III, the wires array can be modelled as a single linear heat source along which no axial temperature gradient occurs at the level of the scanned area.

Figures 11 and 12 were obtained for the array with the 10μm period. They reveal a temperature gradient due to the presence of the cold border of the structure. Contrary to the numerical simulations (see Figure 9) no local heating allows for distinguishing individual wires. A combination of effects related to the measurement technique can explain such results:

(i) As reported in the introduction of this paper, thermal conduction through the air contributes to the heat transfer between the sample and the probe. Its contribution is not negligible and the more so as the scanned sanple is thermal insulating material, as it is the case for our sample. Its radius of action is larger than the solid-solid contact area and the water meniscus width. The corresponding contact radius can be several micrometers due to the size and the geometry of the tip [5]. As a consequence, SThM is sensitive to variations of the heat rate (exchanged between the tip and the sample) averaged on a micrometric area. This may lead to a smoothing of the detected temperature field on the sample surface.

(ii) A local variation of the heat loss from the sample to the tip is detected only if the SThM sensitivity allows for it. We determined the order of magnitude of the sensitivity corresponding to our experimental conditions by use of the following expression [13]:

$$s = \frac{\partial V_{AB\omega}}{\partial T_{\omega}} = E\alpha \frac{1}{Lm} \frac{ch(mL)-1}{ch(mL)-S_p K_p R_{th} msh(mL)} \qquad (7)$$

where $m^2 = j\omega.a^{-1} + h.p.(K_p.S_p)^{-1}$, and $E = CVR_1R_p/(R_1+R_p)^2$ is a function which depends on the electronic components used in the amplification chain stage. $V_{AB\omega}$ and $T_{\omega}$ are respectively the measurement system output voltage and the top sample surface temperature at the modulation pulsation $\omega$. $\alpha$, $a$, $K_p$ $R_{th}$, $S_p$, $L$, $p$ and $h$ are respectively the temperature coefficient of the probe electrical resistance, the thermal diffisivity and the thermal conductivity of the platinum tip, the total thermal tip-sample contact resistance, the section of the probe, the half-length and the circular perimeter of the platinum wire, and the platinum wire convection-radiation coefficient.

We calculated this sensitivity for the frequency of 140Hz. In Table 4 are listed the all used SThM thermal resistive probe properties and acquisition circuit characteristics used in this determination. For the discussed experiment, the sensitivity varies between 10 to 70μV.K$^{-1}$ on the range of the possible values of the thermal resistance of the tip-sample contact: $5.10^4$ - $10^6$ K.W$^{-1}$ [4-6]. The used lock-in amplifier provides a stable signal up 10μV with our bridge. A local temperature increase of about 1K above each resistor is predicted from numerical simulations (see Figure 9) but no thermal resistance between the resistive lines and the passivation layer was taken into account in our model. The effective presence of such thermal resistances confines heat in the sample oxide. Our simulations then certainly overestimate the amplitude of the surface ac temperature field. In these conditions, the different heated lines could not be detected.

A smoothing of the detected temperature field on the sample surface due to a thermal exchange on a micrometric area, a poor thermal conductance between the used probe (of which the apex surface is not well known) and a lack of sensitivity of our set up can then explain the presented results. The limit of detection of the technique would then be reached in our experiments. As a result, the spatial resolution of the technique seems here larger than 10μm.

*B. 'Samples 2': With Thin Passivation Layer*

Figure 13 gives the topography and the thermal contrast obtained with the SThM above one of the nine resistive lines for a sample of type "samples 2". The array was submitted to a 11 mA sinusoidal current at f$_s$ = 110Hz. The contrast of the thermal image clearly reveals the heating of the resistor. But the comparison of topography profiles given by Figure 13.a and Figure 6.b shows that the thermal probe cannot follow the surface due to its large curvature radius.

As illustrated by Figure 14, this leads to surface dilatation phenomena in the topographical image obtained with the thermal probe. The contrast of the thermal image is also affected by this dilatation effect. The heat rate exchanged between the tip and the sample is indeed directly proportional to the thermal contact area. An artifact is then included in the figure 13.b due to a topography-thermal coupling. Note that the dips on both sides of the wires are insulators. If the surface was flat, heat spreading might also hide the real size of the wires.

For this sample, the SThM mounted with the 5μm wire allows to detect submicrometric elements but the artefact due to topography-thermal coupling enlarges the size of the wires by a factor of 10.

## IV. CONCLUSION

AC surface temperature fields of several ICs including arrays of resistive lines with submicrometric sections were investigated with a SThM. The results for flat chips including buried heat sources in one hand and components with apparent heat sources in the other hand highlight the irrelevance of the SThM to handle complex ICs design and topographies, especially due to the size of the thermal probe itself.

When an IC passivation layer is too thick compared to the space between submicrometric integrated heat sources to permit a surface temperature contrast, submicrometric elements resolution will not of course be achieved due to heat spreading in this passivation layer. In the opposite case, the large area of the thermal contact between the probe and the sample, resulting from the thermal conduction through the surrounding gas, may limit the spatial resolution. In order to improve this resolution, one of our further works will concern measurements in vacuum.

The poor resolution obtained in our experiments may also be explained by the lack of sensitivity of our set up. Parameters defining this sensitivity, such as probe-sample thermal conductance and the apparatus for signal generation and processing, limit the minimum measurable heat rate between the probe and the sample and temperature variation. The spatial resolution is then strongly linked with these parameters to be specified.

Finally, we also show that the spatial resolution may also be affected by topography-thermal coupling. The large curvature radius of the probe may particularly lead to large dilatation of the ICs topography and surface temperature.

ICs appear as interesting tools to benchmark the spatial resolution of thermal metrologies such as SThM. The thermal design and the composition of the structure has however to be precisely defined and controlled to make quantitative temperature measurements. Flat samples with various and well known physical properties and dimensions are notably needed to develop an accurate thermal model of the probe-sample system in order to finally achieve the calibration of the technique.

ACKNOWLEDGMENT

The authors are thankful to the laboratory TIMA: Techniques of Informatics and Microelectronics for computer Architecture (Grenoble, France) for supplying the samples.

**First A. Author** (M'76–SM'81–F'87) and the other authors may include biographies at the end of regular papers. Biographies are often not included in conference-related papers. This author became a Member (M) of IEEE in 1976, a Senior Member (SM) in 1981, and a Fellow (F) in 1987. The first paragraph may contain a place and/or date of birth (list place, then date). Next, the author's educational background is listed. The degrees should be listed with type of degree in what field, which institution, city, state or country, and year degree was earned. The author's major field of study should be lower-cased.

The second paragraph uses the pronoun of the person (he or she) and not the author's last name. It lists military and work experience, including summer and fellowship jobs. Job titles are capitalized. The current job must have a location; previous positions may be listed without one. Information concerning previous publications may be included. Try not to list more than three books or published articles. The format for listing publishers of a book within the biography is: title of book (city, state: publisher name, year) similar to a reference. Current and previous research interests ends the paragraph.

The third paragraph begins with the author's title and last name (e.g., Dr. Smith, Prof. Jones, Mr. Kajor, Ms. Hunter). List any memberships in professional societies other than the IEEE. Finally, list any awards and work for IEEE committees and publications. If a photograph is provided, the biography will be indented around it. The photograph is placed at the top left of the biography. Personal hobbies will be deleted from the biography.